\documentclass[conference]{IEEEtran}
\IEEEoverridecommandlockouts

\usepackage{cite}
\usepackage{amsmath,amssymb,amsfonts}
\usepackage{algorithmic}
\usepackage{graphicx}
\usepackage{textcomp}
\usepackage{xcolor}
\usepackage{booktabs}
\usepackage{multirow}
\usepackage{threeparttable}
\usepackage{url}
\usepackage[colorlinks=true, linkcolor=black, citecolor=black, urlcolor=black]{hyperref}

\def\BibTeX{{\rm B\kern-.05em{\sc i\kern-.025em b}\kern-.08em
    T\kern-.1667em\lower.7ex\hbox{E}\kern-.125emX}}

\begin{document}

\title{Efficient Video to Audio Mapper with Visual Scene Detection\\
}

\author{\IEEEauthorblockN{Mingjing Yi}
\IEEEauthorblockA{\textit{Division of Natural and Applied Science} \\
\textit{Duke Kunshan University}\\
Kunshan, China \\
mingjing.yi@dukekunshan.edu.cn}
\and
\IEEEauthorblockN{Ming Li}
\IEEEauthorblockA{\textit{Suzhou Municipal Key Laboratory of Multimodal Intelligent Systems} \\
\textit{Duke Kunshan University}\\
Kunshan, China \\
ming.li369@dukekunshan.edu.cn}
}

\maketitle

\begin{abstract}
Video-to-audio (V2A) generation aims to produce corresponding audio given silent video inputs. This task is particularly challenging due to the cross-modality and sequential nature of the audio-visual features involved. Recent works have made significant progress in bridging the domain gap between video and audio, generating audio that is semantically aligned with the video content. However, a critical limitation of these approaches is their inability to effectively recognize and handle multiple scenes within a video, often leading to suboptimal audio generation in such cases. In this paper, we first reimplement a state-of-the-art V2A model with a slightly modified light-weight architecture, achieving results that outperform the baseline. We then propose an improved V2A model that incorporates a scene detector to address the challenge of switching between multiple visual scenes. Results on VGGSound show that our model can recognize and handle multiple scenes within a video and achieve superior performance against the baseline for both fidelity and relevance.

\end{abstract}

\begin{IEEEkeywords}
audio generation, video to audio, machine learning
\end{IEEEkeywords}

\section{Introduction}
In the era of Artificial Intelligence-Generated Content (AIGC), there has been rapid advancement in various types of generative AI technologies. Most commonly known generative AI systems, such as ChatGPT for language modeling \cite{liu2023summary}, Stable Diffusion for text-to-image generation \cite{rombach2022high}, and AudioLDM for text-to-audio generation \cite{liu2023audioldm}, primarily utilize text as the main prompting modality. 

Video-to-audio (V2A) generation, while not a new topic, presents unique challenges due to its requirement for cross-modal feature transformation. Early studies aiming to bridge the visual and acoustic domains have largely concentrated on image-to-audio generation \cite{chen2018visually,10096023,chen2017deep}. However, video data, conceptualized as sequences of images extended along the temporal dimension, introduces additional complexity. Initial V2A research efforts have typically focused on generating audio within more inclusive domains rather than targeting general sound synthesis \cite{su2020audeo}.

Recent advancements have explored various approaches to improve cross-modal and joint visual-audio generation. For instance, "Seeing and Hearing" \cite{Xing_2024_CVPR} developed an optimization-based framework for integrated visual and audio generation. SpecVQGAN \cite{iashin2021taming} trained a codebook for audio and utilized discrete tokens predicted from video data to synthesize corresponding audio outputs.Diff-Foley \cite{luo2024diff} proposed a way to generate audio from video utilizing latent diffusion model. Wang et al. introduced V2A-Mapper \cite{wang2024v2a}, which leverages multiple foundation models including CLIP \cite{radford2021learning}, CLAP \cite{wu2023large}, and AudioLDM \cite{liu2023audioldm} to generate audio from visual input. CLIP and CLAP employ image-text and audio-text pairs, respectively, to train models capable of extracting semantic-level features from images and audio. These models effectively bridge audio and video within a shared textual space, using the mapper to predict audio features based on the video inputs. Subsequently, these audio features condition on a diffusion model, pre-trained on CLAP-audio pairs, to generate audio that semantically aligns with the video content.

\begin{figure*}[h]\
\centering
\centerline{\includegraphics[width=0.85\textwidth]{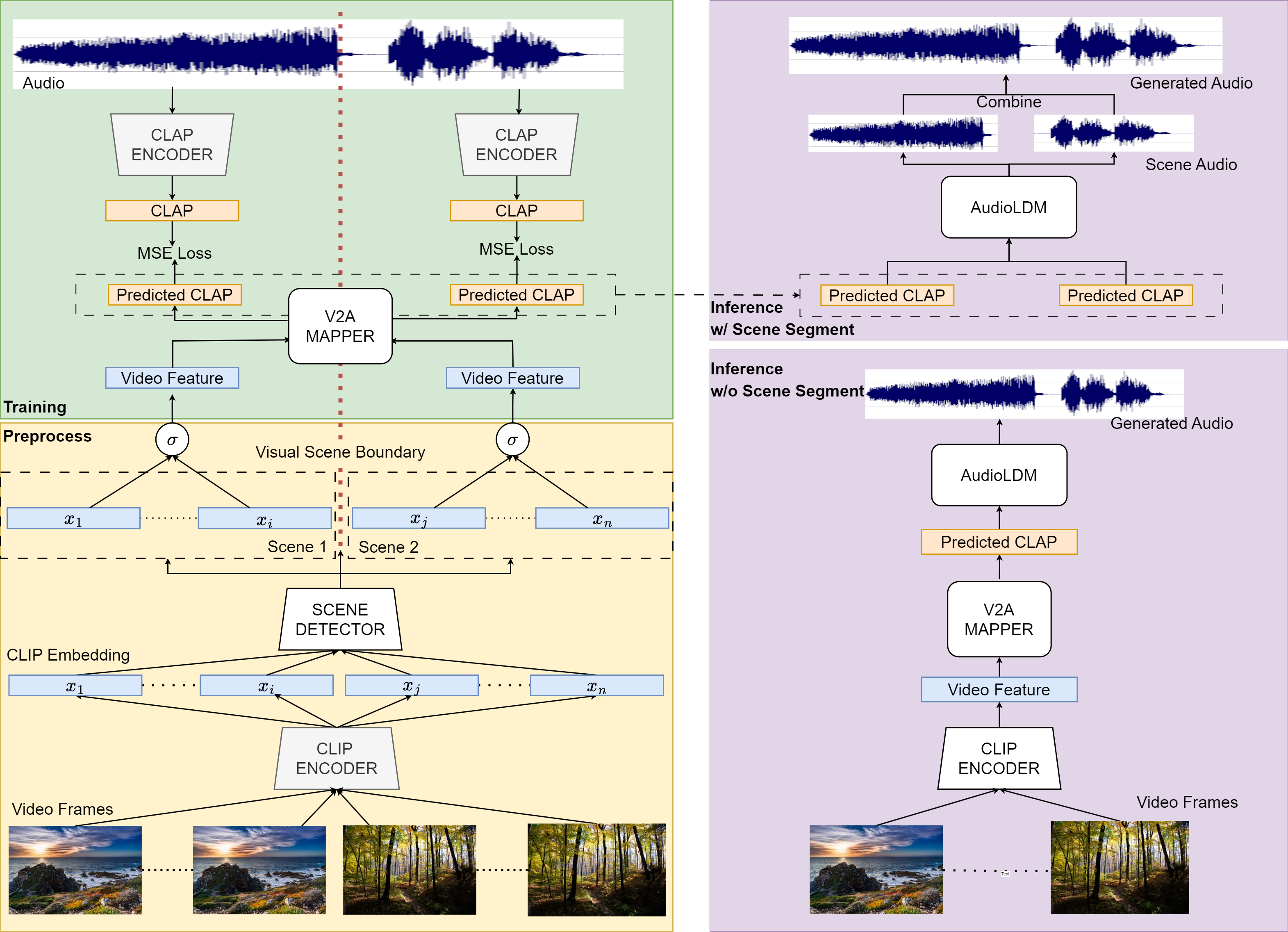}}
\caption{\textbf{Overview of our V2A model. Left: The training process of V2A-SceneDetector.} We utilize pretrained CLIP and CLAP models for feature extraction. By utilizing scene detector, we can identify the scene information and scene boundary between scenes for audio segmentation. \textbf{Right: Inference pipeline.} The top one shows the pipeline with scene segmentation. Sharing the preprocess with training process, we condition on the predicted CLAP embedding to generate audio via AudioLDM. the bottom shows the process of inference without scene segmentation.}
\label{fig1}
\end{figure*}

Despite these advances, existing models exhibit certain limitations. One notable issue is the lack of temporal synchronization between video and audio, even when utterance-level semantic alignment is achieved. Recent efforts have sought to address this challenge. For example, SyncFusion \cite{comunita2024syncfusion} introduced an onset detector and incorporated onset embeddings from videos alongside the CLAP features of audio to enhance synchronization. Another approach by Wang et al. \cite{wang2024frieren} employed a non-autoregressive vector field estimator based on a feed-forward transformer and implemented channel-level cross-modal feature fusion to achieve robust temporal alignment. MaskVAT \cite{pascual2024masked} combined a general audio codec with a sequence-to-sequence generative model, resulting in audio outputs that are both semantically and temporally aligned with the video. FoleyCrafter \cite{zhang2024foleycrafter} proposed a Semantic Adapter (SA) and Temporal Controller (TC), where the TC includes a timestamp detector and a timestamp-based adapter to improve audio-video synchronization.

Another critical issue with current models is their inability to accurately generate audio when the input video contains multiple scenes. This limitation is particularly significant given the growing demand for generating audio for videos with longer durations and multiple scenes. To address this challenge, we employ change point detection to handle audio generation for videos with multiple scenes. Our approach builds upon the framework of Wang et al. \cite{wang2024v2a} with modifications to the mapper architecture. Through the exploration of structures, we found that even relatively simple mapper architectures can achieve satisfactory results. By incorporating scene segmentation techniques to detect and segment scenes within the original dataset, our model produces audio with higher fidelity and better semantic alignment than the baseline models, owing to the increased purity and semantic consistency of the data. The samples and codes for our work are available\footnote{\url{https://1mageyi.github.io/V2A-SceneDetector.demo/}}.

\section{Method}

\subsection{Scene Detector}

We aim to detect scene boundaries in a sequence of frame embeddings using change point detection. Let \(\mathbf{X} = [\mathbf{x}_1, \mathbf{x}_2, \ldots, \mathbf{x}_T] \in \mathbb{R}^{T \times D}\) represent the sequence of frame embeddings, where \(T\) is the number of frames, and \(D\) is the embedding dimension (e.g., 512 for CLIP embeddings).

The self-similarity matrix \(\mathbf{S} \in \mathbb{R}^{T \times T}\) is defined by the inner product between all pairs of embeddings:
\[
\mathbf{S}_{i, j} = \mathbf{x}_i \cdot \mathbf{x}_j = \sum_{k=1}^{D} x_{i, k} \cdot x_{j, k}, \quad \text{for } i, j = 1, 2, \ldots, T.
\]
To enhance the detection of changes, we center the self-similarity matrix by subtracting the mean of rows and columns:
\[
\mathbf{S}_{\text{centered}} = \mathbf{S} - \text{mean}(\mathbf{S}, \text{axis}=0) - \text{mean}(\mathbf{S}, \text{axis}=1)^\top.
\]
Next, we compute the difference between consecutive rows of the centered similarity matrix to capture changes over time:
\[
\Delta \mathbf{S}_{i} = \sum_{j=1}^{T} \left| \mathbf{S}_{i, j} - \mathbf{S}_{i+1, j} \right|, \quad \text{for } i = 1, 2, \ldots, T-1.
\]
This yields a sequence \(\Delta \mathbf{S} = [\Delta \mathbf{S}_1, \Delta \mathbf{S}_2, \ldots, \Delta \mathbf{S}_{T-1}]\) which quantifies the change between consecutive frames.
To identify significant changes, we detect peaks in \(\Delta \mathbf{S}\). Peaks correspond to potential scene boundaries and are identified using a threshold \(\tau\), which scales the standard deviation of \(\Delta \mathbf{S}\):
\[
\text{Peaks} = \{ i \mid \Delta \mathbf{S}_i > \tau \cdot \sigma_{\Delta \mathbf{S}} \},
\]
where \(\sigma_{\Delta \mathbf{S}}\) is the standard deviation of \(\Delta \mathbf{S}\), and \(\tau\) is an adjustable threshold factor.
Visual scene boundaries are determined by the detected peaks. Define the scene boundaries \(\{(s_k, e_k)\}\) such that:
\begin{itemize}
    \item Each scene starts at \(s_k\) (either 0 or right after a peak).
    \item Each scene ends at \(e_k\) (right before the next peak or at the final frame).
\end{itemize}
The final set of scenes is:
\[
\{(s_k, e_k)\} = \{(0, p_1-1), \ldots, (p_{n-1}, p_n-1), (p_n, T-1)\},
\]
where \(p_i\) are the indices of detected peaks in the sequence \(\Delta \mathbf{S}\).
This approach allows for dynamic and adjustable detection of scenes by controlling the sensitivity of peak detection via the threshold \(\tau\).
\subsection{V2A-MLP}

\begin{table}[h]
    \centering
    \caption{Architecture of V2A-MLP}
    \begin{tabular}{c|c|c}
        \hline
        \textbf{Layer} & \textbf{Input Dimension} & \textbf{Output Dimension} \\
        \hline
        Linear & 512 & 1024 \\
        \hline
        ReLU & 1024 & 1024 \\
        \hline
        Linear & 1024 & 512 \\
        \hline
        Linear & 512 & 512 \\
        \hline
        Linear & 512 & 512 \\
        \hline
    \end{tabular}
    
    \label{tab:mlpmodel1}
\end{table}

We adopt the pipeline outlined in \cite{wang2024v2a} to train our baseline model. In the original work \cite{wang2024v2a}, a diffusion-based transformer \cite{vaswani2017attention} is employed to train the mapper. In contrast, our approach utilizes a multi-layer perceptron (MLP) architecture for the mapper. Specifically, we use pre-trained CLAP and CLIP encoders to extract audio and video embeddings from the input videos. Our MLP-based mapper is trained using Mean Squared Error (MSE) loss to align the video features with the corresponding audio features.

\begin{table*}[h]
    \centering
    \caption{Model Performance Metrics}
    \begin{threeparttable}
        \begin{tabular}{ll|ccccc}
            \toprule
            Model & Scene Segment & FD ($\downarrow$) & MKL ($\downarrow$) & CLIP-Score ($\times 10^{-2} \uparrow$) & FAD ($\downarrow$) & LSD ($\downarrow$)\\
            \midrule
            \midrule
            Reference & - & 0 & 0 & 6.047 & 0 & 0\\ 
            Diff-Foly & - & 21.659 & 3.173 & 7.284 & 6.522 & 3.020\\
            V2A-Mapper & - & 11.835 & 2.686 & 8.652 & \textbf{1.029} & 1.456\\
            \cmidrule{1-7}
            \multirow{2}{*}{V2A-MLP (ours)} & w/o segment & 19.0034 & 2.395 & 8.495 & 3.0832 & 1.387\\
            & w segment & 17.366 & 2.401 & 7.632 & 3.524 & 1.368\\
            \cmidrule{1-7}
            \multirow{2}{*}{V2A-SceneDetector (ours)} & w/o segment & 14.6492 & 2.248 & \textbf{9.056} & 2.433 & 1.400\\
            & w/ segment & \textbf{10.742} & \textbf{2.244} & 8.527 & 2.224  & \textbf{1.358}\\
            \bottomrule
        \end{tabular}
        \begin{tablenotes}
            \footnotesize
            \item  For "reference", we evaluate the matric using the real audio to obtain the reference value for comparision.
        \end{tablenotes}
    \end{threeparttable}
\end{table*}

\subsection{V2A-SceneDetector}
To enhance the V2A architecture, we integrate a scene detector, as illustrated in Figure \ref{fig1}. This detector identifies visual scene boundaries by analyzing CLIP embeddings extracted from each frame of the video. Once the scene boundaries are identified, we segment both the audio and CLIP sequences into subparts, ensuring that each segment of the video corresponds to a single scene. The mapper is then trained separately on these segmented and aligned audio and video features, thereby improving the model’s ability to generate scene-specific audio content.

\subsection{Inference}
We propose two inference methods for our V2A model. The first method, without scene segmentation, adheres to the pipeline described in \cite{wang2024v2a}. In this approach, CLIP features are extracted from the input video and aggregated to form a comprehensive video feature. The mapper then predicts the corresponding CLAP embedding, which is subsequently used by AudioLDM to generate audio.

The second method incorporates scene segmentation. Using the scene detector, we identify the scene boundaries within the video and generate audio for each scene independently. The resulting audio segments are then combined into a single audio file that corresponds to the entire video. This approach aims to enhance the temporal alignment and contextual relevance of the generated audio, particularly in videos with multiple scenes.

\section{Experiments}

\subsection{Dataset}
We utilize the VGGSound video dataset \cite{chen2020vggsound} for training and testing. VGGSound comprises 199,176 ten-second video clips sourced from YouTube. Following the data split method from \cite{wang2024v2a}, we use 183,730 videos for training and 15,446 for testing. The original audio sample rate in the dataset is 44,100 Hz; however, we downsample the audio to 16,000 Hz to enhance computational efficiency and ensure consistency in metric comparisons.

\subsection{Metric}
To evaluate the fidelity of the generated audio, we use the Mean KL Divergence (MKL) \cite{iashin2021taming}, Fréchet Distance (FD), and Fréchet Audio Distance (FAD) \cite{kilgour2019frechetaudiodistancemetric} and Log-Spectral Distance (LSD) \cite{mandel2023aero}. MKL measures the divergence between the probability distributions of the features of the generated and real audio. FD measures the distance between the multivariate Gaussians fitted to embeddings of real and generated samples, capturing both mean and covariance differences. Evaluation tools from \cite{liu2023audioldm} are employed for these tasks. LSD is a metric used to evaluate the similarity between two audio signals by measuring the difference between their log-magnitude spectrograms. It helps assess the quality of generated audio by quantifying how closely it matches real audio in the frequency domain, making it useful for evaluating general audio quality and fidelity. Additionally, we use the CLIP Score, as in \cite{wang2024v2a}, to assess the semantic relevance between the generated audio and real audio. Wav2CLIP \cite{wu2022wav2clip} is used to extract CLIP embeddings from the generated audio. We compute the cosine similarity between these embeddings and each frame of the real video, averaging the results to obtain the CLIP score for each audio sample. The mean CLIP score of all generated audio samples is reported as the final result.

\subsection{Experiment settings}
For our experiments, we employ the "ViT-B/32" pretrained checkpoint as the CLIP model \cite{radford2021learning} and the "audioldm-s-full" version of the AudioLDM model \cite{liu2023audioldm} with guidance scale of 4.5. The CLAP encoder \cite{wu2023large} is set to the "htsat-tiny" model to ensure alignment with AudioLDM. The V2A-MLP mapper utilizes the MLP architecture specified in Table \ref{tab:mlpmodel1}
, with mean pooling as the aggregation method, denoted by $\sigma$. We optimize the models using the AdamW optimizer with a learning rate of 0.0001. 

In the V2A-SceneDetector configuration, we set the scene detector threshold to 5 to achieve general scene detection accuracy. Video segments are filtered by validating their duration; segments shorter than 2 seconds are considered invalid and excluded from training. This filtering process yielded approximately 210,000 valid training samples. Both the V2A-MLP and V2A-SceneDetector models are trained for 100 epochs on this dataset.

For testing, we evaluate the models using four variants: V2A-MLP, V2A-SceneDetector, V2A-MLP with scene segmentation, and V2A-SceneDetector with scene segmentation.

\subsection{Results}
We benchmark our models against state-of-the-art (SOTA) V2A systems using samples generated by \cite{wang2024v2a} and test audio produced with Diff-Foley \cite{luo2024diff} from the VGGSound dataset. For the generation process, we set the step parameter to 25 and the guidance scale to 4.5.

Our evaluation reveals that the lightweight V2A-MLP model achieves comparable results to SOTA models in terms of MKL, LSD and FD and improvements in MKL. The integration of the scene detector further enhances performance, with improvements of up to 24\% in relevance (CLIP score) and notable advancements in fidelity. Crucially, our model demonstrates an enhanced ability to handle videos containing multiple scenes. Unlike previous models, which often struggle to generate semantically aligned audio for such videos, our approach significantly reduces alignment errors, showcasing its effectiveness in complex scenarios.

Within our four model variants, we observe notable differences:

\subsubsection{Impact of Using Scene Detector} Comparing models with and without the scene detector, we find that incorporating the scene detector consistently improves performance across all evaluation metrics. The observed improvements range from 6\% to 26\%. This enhancement can be attributed to the model's ability to learn from cleaner video segments that each contain a single scene, thereby improving the fidelity and relevance of the generated audio.

\subsubsection{Impact of Scene Segmentation During Inference} Evaluating the effect of applying scene segmentation at the inference stage, we note that this approach leads to better fidelity as measured by FD, MKL, FAD and LSD scores. However, the CLIP score shows a slight decrease. This may be due to some test videos containing multiple scenes, leading to inconsistencies in the CLIP embeddings that could introduce errors during inference.

\section{Conclusion and Future Work}
In this paper, we explored and extended the capabilities of Video-to-Audio (V2A) generation by integrating scene detection into the V2A framework. Building upon the baseline established by \cite{wang2024v2a}, we introduced two variants of our model: V2A-MLP and V2A-SceneDetector. We raise an approach of visual scene detector, which process the frame embedding sequence using boundary detection. Without external feature extraction, the scene detector cam handle the scene segmentation work at a relatively low cost. We demonstrated that incorporating a scene detector significantly enhances the performance of V2A models by enabling the system to learn from more consistent and semantically coherent video segments. This improvement is reflected in the substantial gains across fidelity metrics such as FD, MKL and FAD. Moreover, we investigated the impact of scene segmentation during inference, finding that it generally boosts the fidelity of the generated audio. More importantly, the integration of scene detector provide the system ability to handle V2A tasks where videos may contain multiple scenes. 

Despite these advancements, the current models still face several limitations. One issue is the variability in the duration of video segments, while AudioLDM can only generate audio in multiples of 2.5 seconds, necessitating the segmentation of redundant parts of generated audio clips. Additionally, transitions between scenes are not always smooth; the model often struggles to recognize closely connected scenes through sound, leading to abrupt audio shifts. Temporal synchronization between video and generated audio remains another significant challenge. Although recent works \cite{comunita2024syncfusion, zhang2024foleycrafter, wang2024frieren} have made strides in addressing this issue, further research is needed to maintain the sequential information from video and generate audio with improved synchronization. 

Future work will focus on refining these aspects, particularly enhancing temporal alignment and addressing smoothness in scene transitions. Furthermore, we will conduct comprehensive comparisons with state-of-the-art methods to validate the robustness and generalizability of our approach. Overall, our results underscore the potential of scene-aware V2A models in producing high-fidelity, semantically relevant audio that closely aligns with the visual content, paving the way for more sophisticated audio generation techniques in complex video environments.

\bibliographystyle{ieeetr}
\bibliography{reference}

\end{document}